# Design and Analysis of a Detuned Series-Series IPT System with Solenoid Coil Structure for Drone Charging Applications

Elias Nadi
Department of Electrical and Computer Engineerng
Rowan University
Glassboro, NJ, USA
nadie137@students.rowan.edu

Hua Zhang
Department of Electrical and Computer Engineerng
Rowan University
Glassboro, NJ, USA
zhangh@rowan.edu

*Abstract*—This paper proposes a new coil configuration that uses solenoid ferrites on the receiver side instead of planar ferrite coils that are employed in existing wireless charging systems. The solenoid ferrites are used in the drone legs to help mount the receiver on a moving truck, while the two parts of the transmitter are placed on the truck. To validate this idea, a detuned transmitter series-series (SS) compensated IPT prototype system has been investigated to charge the onboard battery of a drone. The proposed detuned design helps to limit the system output power and transmitter side current at even zero coupling condition and tolerates misalignment positions. Experimental results demonstrate that the system's switching frequency is 245 kHz, and the primary side resonates at 223.5 kHz, which is a detuned design. The transmitting current is limited to 4.5 A at zero coupling condition with a low power loss of only 4.2 W at zero coupling condition. The target achievable power of the system is 50 W with a 10 mm air gap and horizontal and vertical misalignment in the range of [0, 50 mm]. This system has been validated by experiments to charge an 11.1V battery at 48.2W with 88.2% efficiency.

## I. Introduction

Inductive power transfer (IPT) is a promising solution for charging vehicles, robots, and drones, as it provides a convenient and safe charging method that eliminates the need for additional hardware of cables and connectors [1]. In existing designs, efficient charging stations and reliable communication systems are essential for the successful operation of drones [2].

In real applications, the usage of charging stations can completely eliminate the need for manual battery charging of quadrotor UAVs [3]. Specifically, Efficient power transfer in an IPT system requires a compensation circuit to resonate with both the transmitting and receiving coils to provide sufficient voltage and current [4].

The series-series (SS) IPT system, which has a constant current (CC) output, is a commonly used solution due to its high circuit simplicity and good compatibility with battery charging. However, the proper alignment between coils is important to avoid overpower and overcurrent when the magnetic coupling is weak [5]. In a zero-coupling scenario between the transmitter and receiver, a series LC resonant short-circuit occurs on the transmitter side, causing a significantly high current that can easily damage the system, which is a practical drawback of the series resonant compensation. To address this issue, a high-order LCC-LCC IPT circuit has been proposed [6].

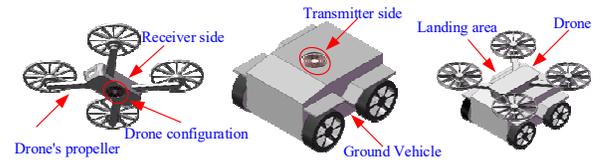

**Fig 1.** The proposed conceptual system configuration.

However, this topology is a high complexity circuit, and its size makes it less attractive. Especially when additional compensation inductor is connected in the system, it is not practical for drone charging applications. High misalignment tolerance is a key point in inductive charging scenarios [7]. Control strategies have been proposed to operate in high misalignment, but these strategies are complicated and rely on real-time communication between the transmitter and receiver sides, which is interfered by magnetic coupler and magnetic field during the charging process [8]-[9].

In this paper, a new configuration for the coils has been proposed. The ferrite for transmitter side is cylindrical and the receiver side is solenoid. A conceptual system configuration for the drone has been extracted based on Fig. 1. The transmitter and receiver coils and ferities each are made of two pieces that are connected in series.

A detuned SS-compensated circuit is adopted. The detuned design has a large inductive impedance, which acts acceptable at weak and even zero coupling conditions. The drone legs are replaced by ferrites, and the transmitter ferrites are installed on the surface of a ground vehicle. It also has good misalignment tolerance compared to existing conventional resonant SS IPT systems. To validate the proposed design, a detuned SS-IPT prototype is implemented with an input dc voltage of 29V and a 11.1V lithium-ion battery load. The experimental results prove that the system limits the input current when coupling is zero and high misalignment problems have been resolved.

## II. Proposed Solenoid Magnetic Coupler Structure Design

In this section, the structure of the transmitter and receiver coils are discussed. At the transmitter side, two pieces of cylindrical ferrites are connected in series to form a primary inductor. Besides, two pieces of solenoid ferrites are connected in series at the secondary side to form the secondary side inductance, as shown in Fig. 2. This model is further used in Maxwell software to simulate parameters.

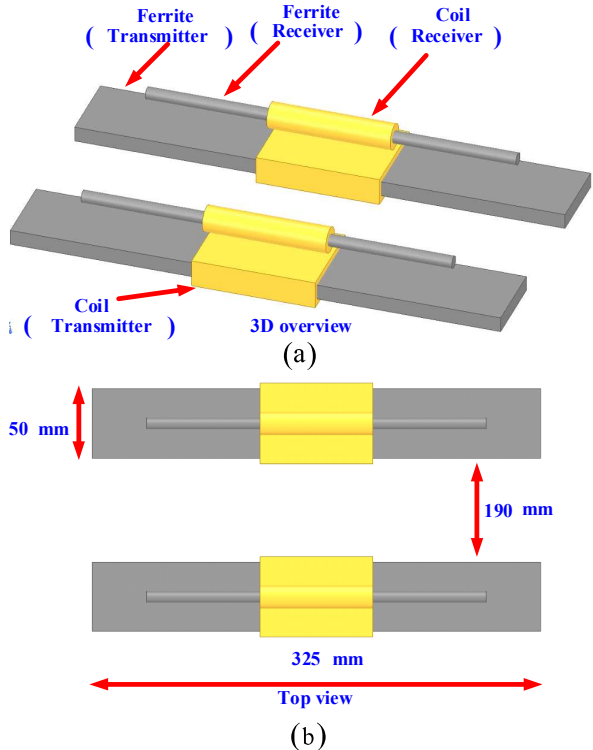

**Fig 2.** The proposed solenoid magnetic coupler structure for drone charging applications.

The transmitter ferrite has been put on the ground vehicle, so the drone will land on it in order to be charged. The solenoid ferrites are used in series as the drone legs. The ferrite length and width for both solenoid ferrites and cylindrical ferrites are optimized in the Maxwell simulation.

The configuration of the inductances is simulated in Maxwell based on the expected dimensions. Figs. 3 and 4 have been extracted from the simulation results for defined value of width and length for both transmitter and receiver and also for 10 mm air gap in the proposed system.

Fig. 3 shows that the coupling coefficient for the proposed system decreases when misalignment in X and Y axis increases, especially in X direction which affects the output power in SS-compensated circuits.

It is shown that for this system when there is no misalignment between the ferrites the coupling coefficient is 0.38. When there is 10 mm misalignment in X axis and 50 mm in Y the coupling coefficient decreases to 26%. The simulation results are consistent with the common sense understanding of the coupling coefficient.

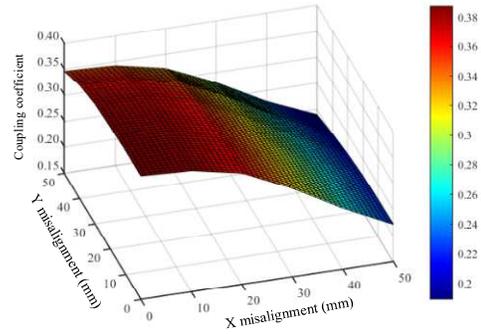

**Fig .3.** Simulated coupling coefficient *vs.* misalignment.

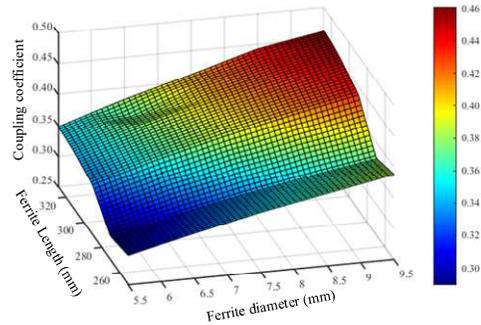

**Fig .4.** Simulated coupling coefficient *vs.* ferrite diameter and length.

Fig. 4 demonstrates the length of ferrite and solenoid ferrite diameter versus different coupling coefficients. Based on this figure, when the diameter and length of solenoid ferrite increases, the coupling coefficient between transmitter and receiver increases as well.

For 328 mm length of the solenoid ferrite and also 8.5mm ferrite diameter, the coupling coefficient 0.38 is achieved. Based on these two figures, the optimal length of ferrites and diameter for receiver side are derived. The distance between two pieces of ferrites is also derived based on the distance between drone legs.

## III. ANALYSIS OF THE PROPOSED DETUNED SS-IPT SYSTEM

The proposed SS-compensated IPT circuit is shown in Fig. 5, including the key parameter values.

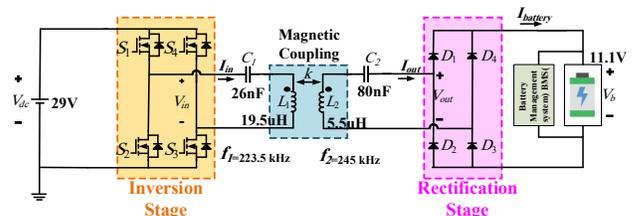

**Fig .5.** The proposed drone charging IPT system and circuit configuration.

In Fig. 5, $L_1$ and $L_2$ represent the transmitter and receiver coils. $C_1$ and $C_2$ are series compensation capacitors. $k$ is the coupling coefficient, and the mutual inductance is represented by $M$. $V_{dc}$ is input DC voltage and $V_b$ is the voltage of the battery load. The system operating switching and angular frequency are defined as $f_s$ and $\omega_s$. The resonant frequency for both transmitter and receiver side are defined as $f_1$ and $f_2$. In a conventional SS compensated system, both transmitter and receiver operate in resonant conditions, thus the switching frequency $f_s$ satisfies. System parameters which are calculated based on the equivalent impedance circuit of SS IPT.

However, working in this condition induces three drawbacks. First, when misalignment occurs, the mutual inductance $M$ reduces, then system current and power increase, this can result in overcurrent and overpower at a weak coupling coefficient which can damage the system.

Second, a conventional SS-compensated system cannot work in an open circuit load condition, because a high voltage will be induced at the secondary side that may damage the system. Third, when $M=0$, the transmitter side forms a short-circuit in the primary LC, which is in a resonance condition, and it will induce a high fault current in the transmitter coil.

This means that the circuit cannot operate alone. To address these problems of a conventional SS system, the primary side should be detuned, and secondary side can work in resonance for simplification. Detuning the primary side can effectively limit the system current at zero coupling condition by controlling the impedance seeing from input side and determines the maximum power of the IPT system, which results in overcurrent and over voltage protection.

In series-series compensated system a battery has been put in the output of the system as the load. The output voltage is hard to regulate and if over voltage happens the battery is broken. In order to regulate the output voltage of the battery, a battery management system (BMS) is proposed to control the voltage and fix it on the nominal voltage to protect it. The output of the system is constant current and by implementing the battery management system, the voltage will also be regulated on the desired voltage level.

## IV. EXPERIMENTAL RESULTS

Fig. 6 shows the implemented drone charging prototype. Table I shows the implemented IPT system parameters. With $V_{dc}$=29 V, $f_s$=245 kHz, and a 3-series (3S) 4500 mAh lithium battery load with nominal voltage of 11.1 V, the performance of the implemented IPT prototype has been investigated through experiments. A silicon carbide (SiC) inverter is designed with the C3M0016120K switches to generate a 245 kHz excitation. A Schottky diode DS6K48003BS low voltage rectifier is also constructed.

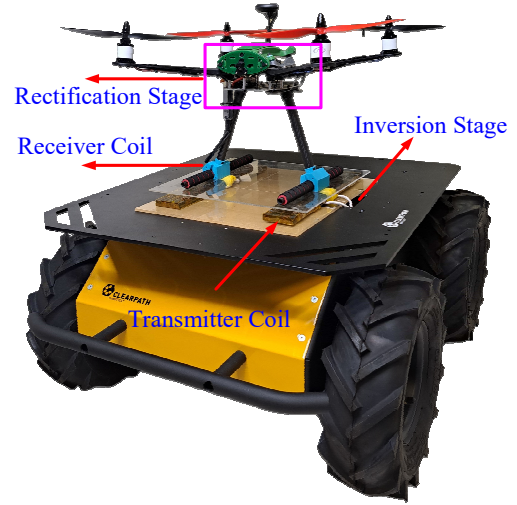

Fig. 6. The implemented charging hardware prototype.

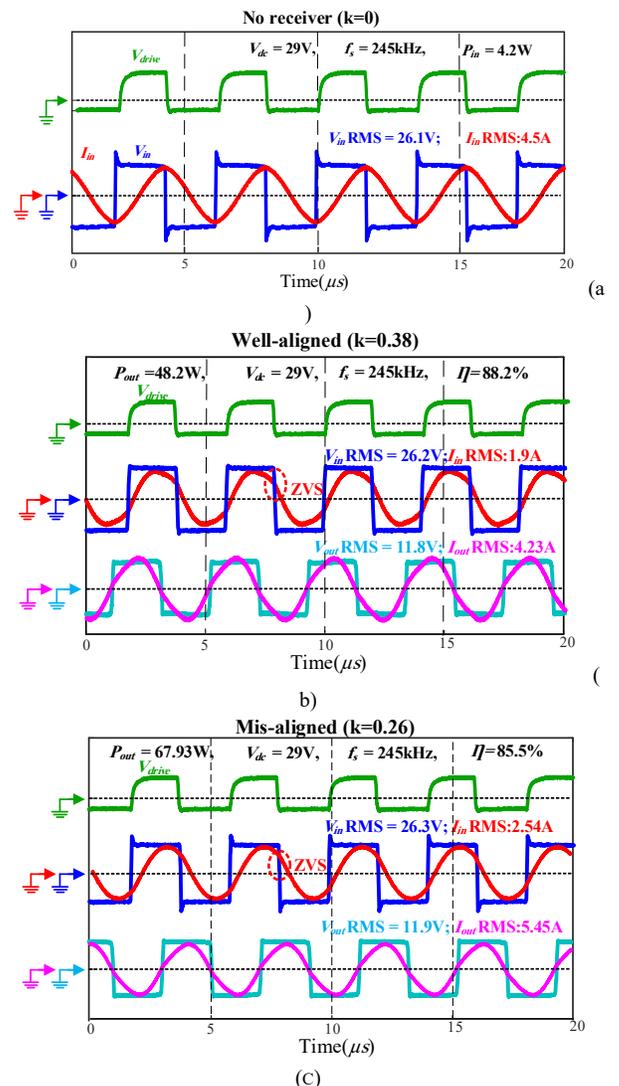

Fig. 7. Experimental waveforms (a) no receiver (b) well-aligned $k$=0.38 (c) 50-mm misalignment $k$=0.26.

Table 1: The implemented IPT System specifications.

| Parameter | Value | Parameter | Value |
|---|---|---|---|
| Input $V_{dc}$ | 29V | $f_1$ | 223.5kHz |
| $S_1$-$S_4$ | C3M0016120K | Distance | 10mm |
| $L_2$ | 5.5uH | $L_1$ | 19.5uH |
| $f_s$ | 245kHz | $C_1$ | 26nf |
| $k$ (well aligned) | 0.38 | $C_2$ | 80nf |
| Load battery | 11.1V | Rectifier | DS6K48003BS |

The new magnetic coupler has a 0.38 coupling coefficient, when the air gap distance is 10 mm. In this system, the operating switching frequency is 245 kHz, and the primary side resonant frequency $f_1$ is set as 223.5 kHz to provide a detuned condition. When the system is at $k=0$, the system power loss only achieves 4.2 W in Fig. 7(a). The input voltage and current of the system are also measured.

It validates that the proposed system can work at the no load condition without damaging the whole system. In the well-aligned situation, the output power is 48.2 W, and the experimental waveforms are shown in Fig. 7(b).

ZVS is achieved in this situation, with Input $V_{dc}$ of 29 V and coupling coefficient of 0.38. When the system is in the 50mm misaligned condition in Y axis and 10mm misalignment in X axis, the output power is 67.93 W, with also coupling coefficients of 0.26. ZVS is also achieved in this situation, which is shown in Fig. 7(c).

## V. Conclusions

This paper proposed a novel magnetic coupler structure and its detuning method for drone charging applications. Instead of using planar coils, solenoid ferrites are adopted. The coils in both transmitter and receiver sides have two parts that are connected in series. To validate this structure, a detuned SS-IPT system are designed. The primary side is in a detuned condition, which limits the transmitting current at zero coupling and misaligned conditions. A 29 V and 245 kHz prototype is implemented with primary side resonant frequency of 223.5 kHz. This hardware is tested in both well aligned and misaligned conditions. Experimental results demonstrate that the transmitting current is 4.5 A at zero coupling condition with a low power loss of 4.2 W. The maximum achievable power of the system is 48.2 W with a 10 mm air gap distance and horizontal and vertical misalignment range of [0, 50 mm]. When the system is in the 50 mm misaligned condition in Y axis and 10 mm misalignment in X axis, the output power is 67.93 W, and ZVS is also achieved in this situation.

## References

[1] S. Wu, C. Cai, A. Wang, Z. Qin and S. Yang, "Design and implementation of a uniform power and stable efficiency wireless charging system for autonomous underwater vehicles," IEEE Trans. Ind. Electron., vol. 70, no. 6, pp. 5674-5684, June 2023.

[2] S. S. Rad et al., "Experimental Study of Magnetic Field Effect on Multi-Rotor Drones Operation in Autonomous Power Line Inspection," 2023 IEEE Transportation Electrification Conference & Expo (ITEC), Detroit, MI, USA, 2023, pp. 1-5.Y.

[3] C. H. Choi, H. J. Jang, S. G. Lim, H. C. Lim, S. H. Cho and I. Gaponov, "Automatic wireless drone charging station creating essential environment for continuous drone operation," 2016 International Conference on Control, Automation and Information Sciences (ICCAIS), Ansan, Korea (South), 2016, pp. 132-136, doi: 10.1109/ICCAIS.2016.7822448.

[4] M. Behnamfar, M. Tariq and A. I. Sarwat, "Novel Autonomous Self-aligning Wireless Power Transfer for Improving Misalignment," in IEEE Access, doi: 10.1109/ACCESS.2024.3374778

[5] Y. Wang, H. Zhang, Y. Cao and F. Lu, "Remaining Opportunities in Capacitive Power Transfer Based on Duality with In-ductive Power Transfer," IEEE Transactions on Transportation Electrification, vol. 9, no. 2, pp. 2902-2915, June 2023, doi: 10.1109/TTE.2022.3225578.

[6] S. Li, W. Li, J. Deng, et al., "A double-sided LCC compensation network and its tuning method for wireless power transfer," IEEE Trans. Veh. Technol., vol. 64, no. 6, pp. 2261-2273, June 2015.

[7] Q. Zhu, Y. Guo, L. Wang, C. Liao and F. Li, "Improving the Misalignment Tolerance of Wireless Charging System by Opti-mizing the Compensate Capacitor," IEEE Trans. Ind. Electron., vol. 62, no. 8, pp.4832-4836, Aug. 2015.

[8] A. Mostafa et al., "Output power regulation of a series-series inductive power transfer system based on hybrid voltage and frequency tuning method for electric vehicle charging," IEEE Trans. Ind. Electron., vol. 69, no. 10, pp. 9927-9937, Oct. 2022.

[9] Y. Wang et al., Compact Z-Impedance compensation for inductive power transfer and its capacitance tuning method," *IEEE Trans. Ind. Electron*, vol. 70, no. 4, pp. 3627-3640, April 2023.